\documentclass[12pt]{iopart}
\usepackage{graphicx}

\def \pt	{p_{\perp}}
\def \pttrig	{p_{\perp}^{\rm trig}}
\def \mpt 	{\langle\pt\rangle}
\def \dphi	{\Delta\phi}

\begin{document}

\title[In-Medium Properties of Jets]{In-Medium Properties of Jets}

\author{F Wang}

\address{Department of Physics, Purdue University, West Lafayette, Indiana 47907, USA}
\ead{fqwang@purdue.edu}

\begin{abstract}
Modifications of jet-like azimuthal correlations have revealed novel properties of the medium created in relativistic heavy-ion collisions. Experimental results on jet-like 2- and 3-particle correlations, specificly ``punch-through" at high transverse momentum, broadening at low and modest transverse momentum, and particularly the possible experimental evidence for conical flow, are reviewed. Future prospects of jet-like correlations and their physics potential are discussed.
\end{abstract}


\vspace{2pc}

\submitto{\JPG}

\section{Introduction}

Jets and jet-like correlations are good probes to study the medium created in relativistic heavy-ion collisions because their properties in vacuum can be calculated by perturbative quantum chromodynamics. Modifications to their properties in nuclear medium can be used to study the nature of the medium~\cite{Eloss}. While exclusive jet reconstruction is difficult in central heavy-ion collisions at RHIC, jet-like azimuthal correlations with a high transverse momentum ($\pt$) trigger particle have proven to be a powerful alternative~\cite{wp}.

The past a few years have seen much progress in the study of jet-like correlations at RHIC. In this talk, experimental results on jet-like azimuthal correlations from RHIC are reviewed. The review is concentrated on results at mid-rapidity where the highest energy density is achieved, on the away side of trigger particles where partons have to traverse the longest distance leading to the strongest modification, and on charged hadrons which have the largest statistics and thus offer the most detailed information. Three major results are discussed: (i) ``punch-through" at high $\pt$, (ii) away-side broadening at low and modest $\pt$, and (iii) collective medium response to parton energy loss.

\section{``Punch-through" at high $\pt$}

\begin{figure}[hbt]
\begin{minipage}{0.4\textwidth}
\hspace*{0.3in}\includegraphics[width=1.2\textwidth]{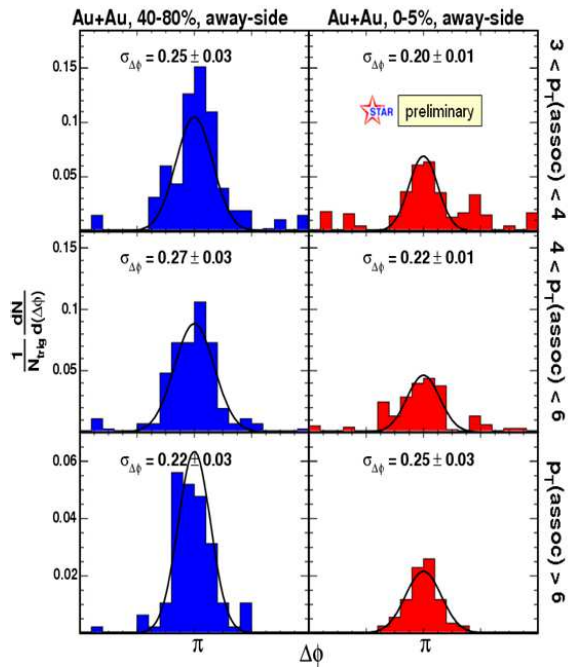}
\end{minipage}
\begin{minipage}{0.6\textwidth}
\caption{(color online) Away-side $\dphi$ correlation peaks in peripheral 40-80\% (left column) and central 0-5\% (right column) Au+Au collisions~\cite{Magestro_paper}. The curves are Gaussion fits to the data. The trigger $\pt$ range is $8<\pttrig<15$~GeV/$c$ and three associated $\pt$ ranges are shown: $3<\pt<4$~GeV/$c$ (upper row), $4<\pt<6$~GeV/$c$ (middle row), and $\pt>6$~GeV/$c$ (lower row).}
\label{fig:gaus_shape}
\end{minipage}
\end{figure}

When the jet energy is sufficiently large, one expects that it would punch through the medium on the away side of the trigger particle. Both STAR~\cite{Magestro_paper} and PHENIX~\cite{Jia_proceedings} have analyzed jet-correlations with large trigger and associated $\pt$, and observed clear away-side correlation peaks charateristic of back-to-back di-jets. Little modification is observed on the near side and the away-side correlation strength is increasingly suppressed with centrality. The shape of the away-side correlation, however, is little changed. Figure~\ref{fig:gaus_shape} shows the away-side correlation peaks together with Gaussian fits from STAR for trigger $8<\pttrig<15$~GeV/$c$ and three associated $\pt$ ranges~\cite{Magestro_paper}. The away-side widths are all similar, for different centralities and $\pt$'s. 
Two possible physics scenarios come to mind: 
(i) There exists a finite probability for the away-side parton not to interact with the medium and fragment in vacuum. The away-side correlation shapes remain the same from peripheral to central collisions; the correlation strength, normalized per near-side jet, is reduced because only a fraction of the away-side partons made out without interaction. This scenario includes the case of tangential di-jets.
(ii) Medium interactions lead to energy loss, either partonic, or both partonic and hadronic~\cite{hadronic_dedx}. After energy loss, the fragmenting parton, now with smaller energy, resemble those triggered by lower $\pttrig$ in peripheral collisions. Indeed, the away-side correlation shape in central collisions is similar to that at lower $\pttrig$ in peripheral collisions~\cite{Magestro_paper}. The underlying physics may be more complex, for example, fragmentation in medium may differ from that in vacuum. The hadron fragments may also lose energy in medium, without significant broadening at large $\pt$. The hadrons, from fragmentation of similar energy partons, may distribute similarly in peripheral and central collisions, but appear at lower $\pt$ in central collisions. 

In order to discriminate these scenarios, one needs additional information. One idea is to trigger on high $\pt$ di-hadrons, and study the correlation of a third, lower $\pt$ hadron. Scenario (i) would yield the same correlation structure for peripheral and central collisions, while scenario (ii) would result in different correlations because the lost energy should appear in soft particles. 

\section{Broadening at Low and Modest $\pt$}

Azimuthal di-hadron correlation studies by STAR~\cite{b2b,jetspectra} have shown that the away-side correlation at modest $\pt$ is strongly suppressed, while the near-side correlation is unchanged. The results demonstrate that the measured high $\pt$ particles stem predominantly from the surface of the collision zone; those jets that are initially produced but have to traverse the medium strongly interact with the medium, and few of their remnants survive at high $\pt$. Preliminary data from PHENIX~\cite{PHENIXb2b} indicate that the away-side correlation in Cu+Cu collisions is less suppressed than in Au+Au, consistent with the smaller medium size in Cu+Cu collisions. 

\begin{figure}[hbt]
\hspace*{1.3in}
\includegraphics[width=0.7\textwidth]{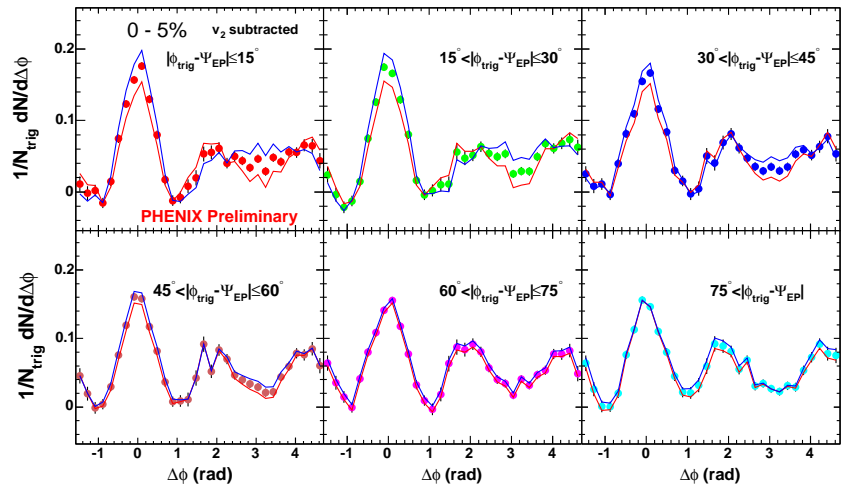}
\vspace{-0.1in}
\caption{(color online) Jet-like azimuthal correlations from PHENIX in top 5\% Au+Au collisions for six orientations of trigger particles relative to the reaction plane~\cite{Jia_proceedings}. The associated and trigger $\pt$ ranges are $1<\pt<2.5<\pttrig<4$~GeV/$c$}
\label{fig:PHENIXb2bRP}
\end{figure}

STAR has further studied the pathlength dependence of away-side suppression via di-hadron correlations with respect to reaction plane~\cite{b2bRP}. The result shows a hint of difference: the away-side correlation out-of-plane is more suppressed than that in-plane, while the near-side correlations are identical, consistent with surface bias and the longer pathlength the away-side parton has to traverse out-of-plane. 
PHENIX has performed a more detailed analysis of di-hadron correlations in top 5\% Au+Au collisions in six bins of the trigger particle orientation from reaction plane~\cite{Jia_proceedings}. The data, depicted in Fig.~\ref{fig:PHENIXb2bRP}, appear to show a systematic trend of increasing away-side suppression from in-plane towards out-of-plane, again consistent with the pathlength dependence of energy loss.

Moving to low $\pt$, STAR has shown~\cite{jetspectra} that the away-side correlated hadrons are broadly distributed and their energy distribution is not much different from that of the bulk medium, indicating partial equilibration.
It has further been shown~\cite{WangQM04,STARmeanpt} that the associated $\mpt$ is the lowest at $\dphi=\pi$, contrary to what is expected from jet fragmentation in vacuum as observed in $pp$ and d+Au. Qualitatively similar results are found by PHENIX~\cite{Jia_proceedings} where the away-side $\pt$ spectrum is softer for the more collimated particles at $\dphi\sim\pi$ in central collisions, and the opposite for peripheral collisions.

\begin{figure}[hbt]
\includegraphics[width=0.32\textwidth,height=0.25\textwidth,bbllx=130,bblly=160,bburx=455,bbury=375]{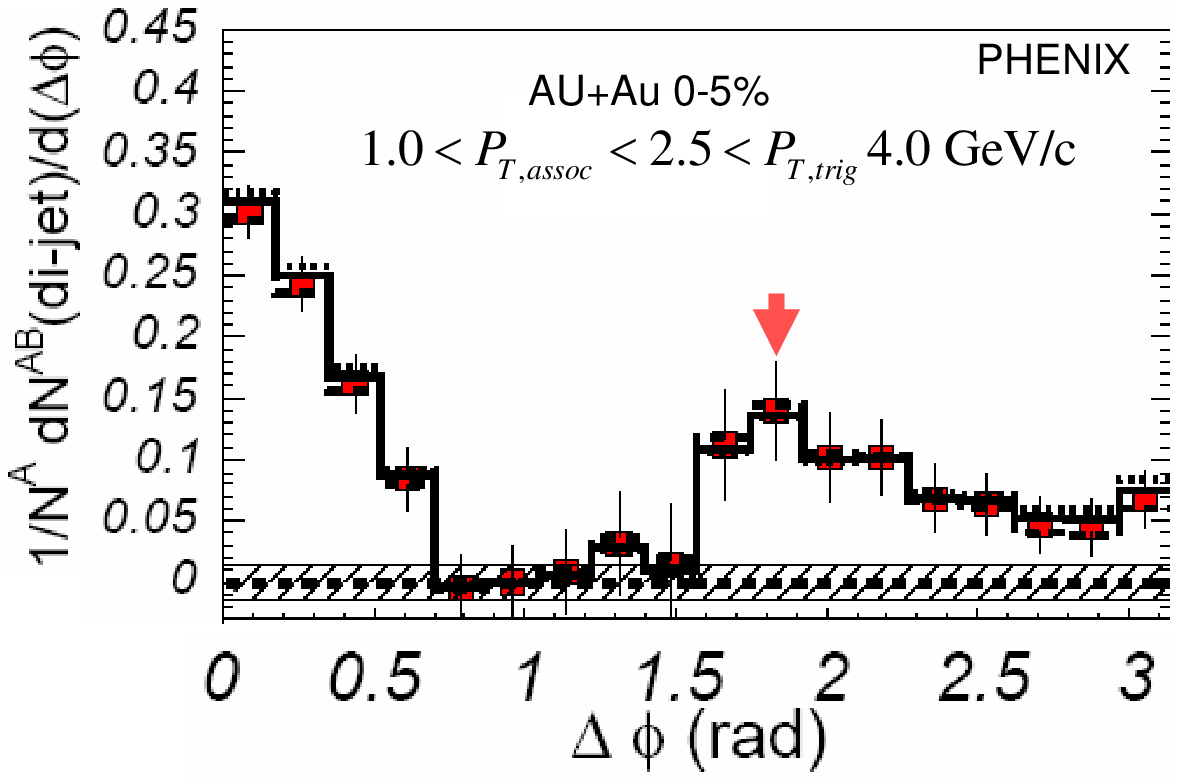}
\includegraphics[width=0.31\textwidth,height=0.238\textwidth,bbllx=20,bblly=0,bburx=565,bbury=400]{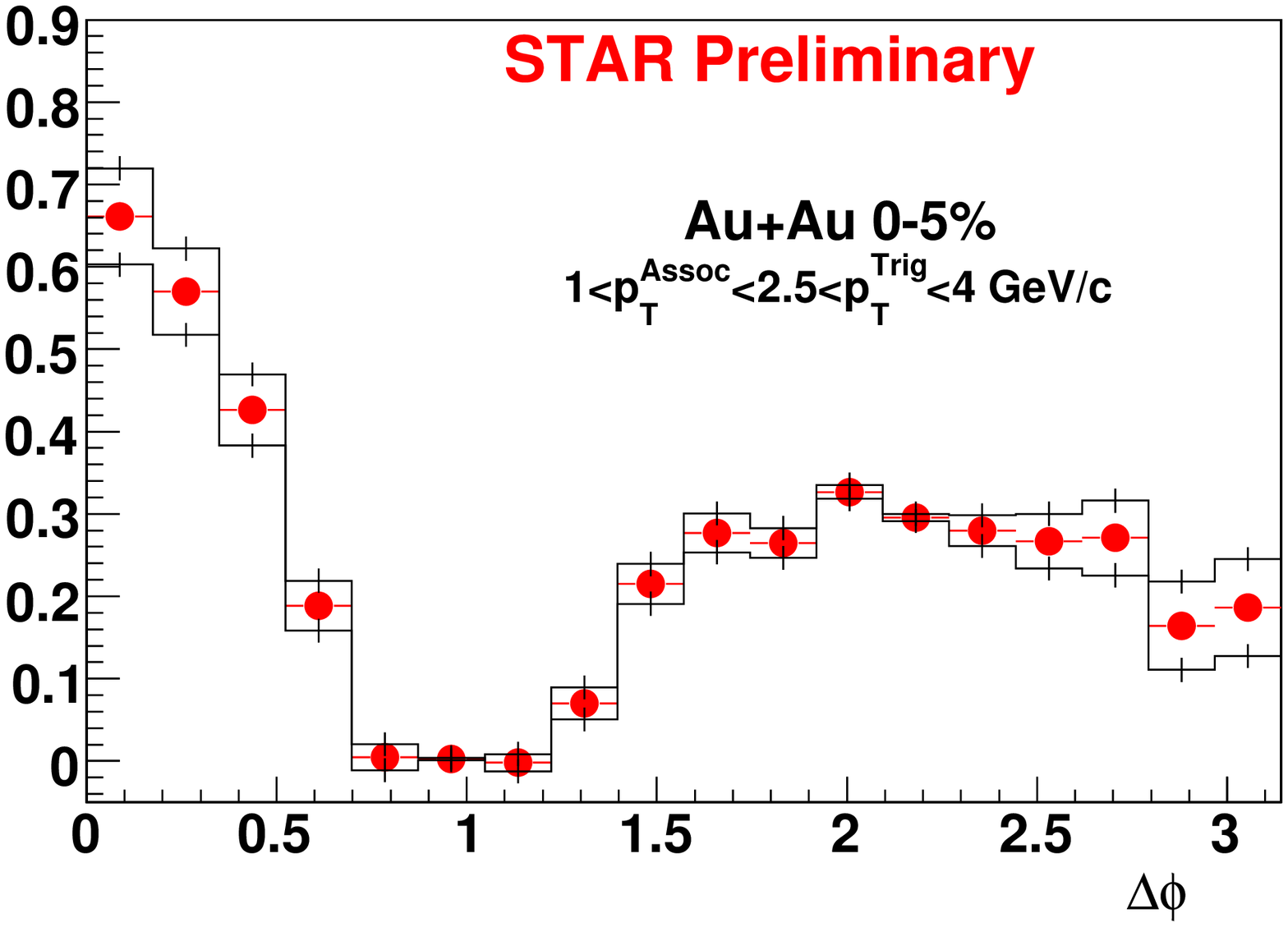}
\includegraphics[width=0.35\textwidth,height=0.24\textwidth,bbllx=0,bblly=5,bburx=510,bbury=395]{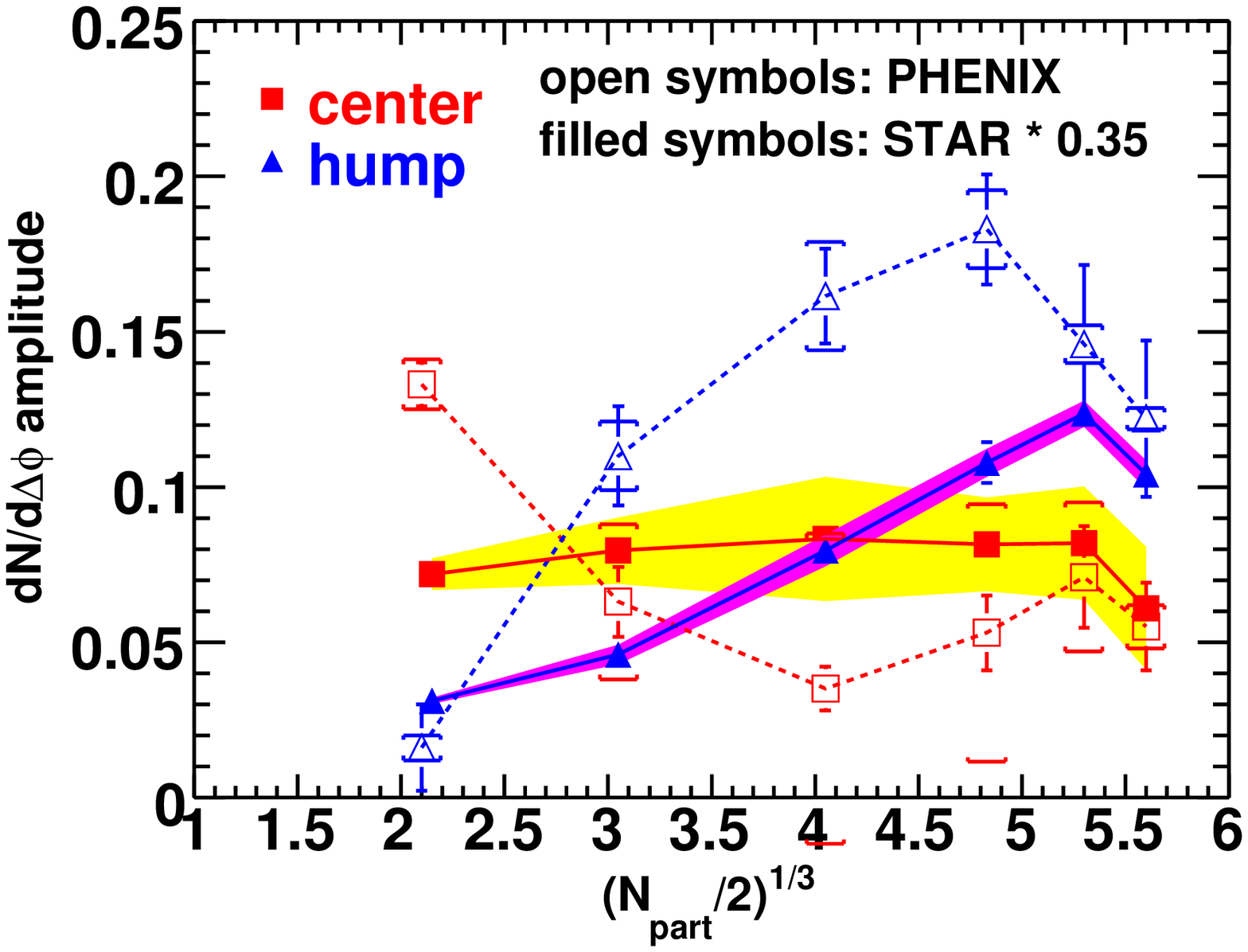}
\caption{(color online) $\dphi$ correlations in top 5\% Au+Au collisions from PHENIX~\cite{dipPHENIX} (left) and STAR~\cite{UleryQM05} (middle). The histograms indicate systematic uncertainties due to flow subtraction. Right: The away-side correlation amplitudes in the central (red) and hump (blue) regions as a function of $(N_{\rm part}/2)^{1/3}$ (where $N_{\rm part}$ is the number of participants). The systematic uncertainties due to flow subtraction are shown in shaded areas for STAR and caps for PHENIX.}
\label{fig:doublepeak}
\end{figure}

The $\mpt$ result implies that the broad away-side distribution becomes broader with increasing associated $\pt$. 
Indeed, this is observed by PHENIX~\cite{dipPHENIX} and STAR~\cite{STARmeanpt,UleryQM05}, depicted in Fig.~\ref{fig:doublepeak} for central Au+Au collisions, in the fixed associated and trigger $\pt$ ranges of $1<\pt<2.5<\pttrig<4$~GeV/$c$; the away-side correlation is even double-humped. The centrality evolution of the double-hump structure is shown in the right panel of Fig.~\ref{fig:doublepeak}, where the correlation amplitudes in the central ($|\dphi-\pi|<\pi/9$) and the hump ($\pi/3<|\dphi-\pi|<4\pi/9$) region are plotted against $(N_{\rm part}/2)^{1/3}$. The STAR data are scaled by the acceptance factor of 0.35. The quantitative discrepancy between the PHENIX and STAR data may come from systematics in flow subtraction. However, the qualitative features are similar: the away-side central amplitude does not drop -- it stays approximately constant -- while the hump amplitude increases with centrality. 

\section{Collective Medium Response and 3-Particle Correlations}

The away-side jet-like correlations are broad. For particular selection of kinematics, the away-side correlation is even double-humped; more associated particles are found in the $\dphi$ regions away from $\pi$ in more central collisions, and those particles are harder.
These results have stimulated many theoretical investigations. 
In particular, Mach-cone shock waves have been suggested as a possible physics mechanism~\cite{machcone1,machcone,machcone2} -- particles are emitted on a cone due to collective excitation; the additional emission of particles results in an increased multiplicity in the hump region, and the shock wave push along the cone direction results in a larger $\mpt$. The generation of Mach-cone shock waves seems inevitable given that jets are supersonic, the medium is hydrodynamic~\cite{wp}, and jets and medium are strongly interacting~\cite{wp}. The Mach-cone angle is determined by the speed of sound in the medium and is independent of the associated $\pt$. Recently, $\check{\rm C}$erenkov gluon radiation is suggested as an alternative mechanism for conical emission~\cite{Cerenkov}; the cone angle in this case is dependent on $\pt$. 

The away-side double-hump structure is consistent not only with conical emission, but also with other physics mechanisms such as large angle gluon radiation \cite{gluonrad} and jet ``deflection'' due to radial flow or preferential selection of particles by the pathlength dependent energy loss mechanism~\cite{Hwa}. In order to discriminate conical emission from other mechanisms, 3-particle azimuthal correlation is needed.


Both PHENIX and STAR have performed 3-particle correlation analysis. The analyses treat the event as composed of two parts: one directly correlated with the trigger particle (``di-jet"), and the other not directly correlated (background). The background is {\em indirectly} correlated with the trigger via the reaction plane. The background is normalized to the 2-particle correlation signal by the common practice of ZYA1 or ZYAM (Zero Yield At 1 radian or Minimum). Two combinatorial backgrounds are present in 3-particle correlation with a trigger particle: pairs of background particles and pairs of a jet-like correlated particle and a background particle~\cite{method}. Careful construction of the backgrounds is critical in 3-particle correlation analysis.


\begin{figure}[hbt]
\hspace{1in}
\includegraphics[width=0.8\textwidth]{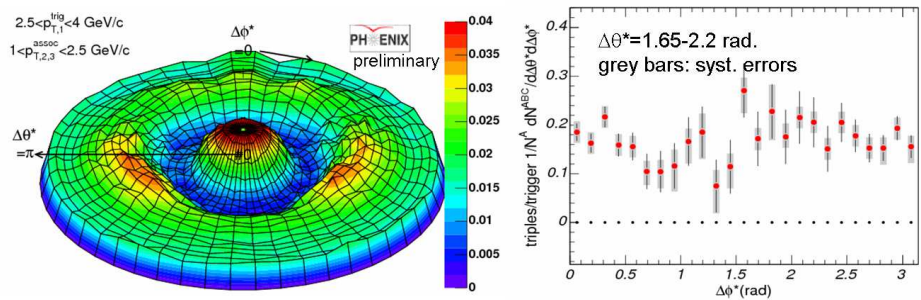}
\vspace{-0.1in}
\caption{(color online) Background subtracted 3-particle correlation in polar coordinate system (left) and its projected $\dphi^*$ distribution (right) from PHENIX for $1<\pt<2.5<\pttrig<4$~GeV/$c$ in 10-20\% Au+Au collisions~\cite{PHENIX3particle}. 
}
\label{fig:phenix_3-particle}
\end{figure}

PHENIX~\cite{PHENIX3particle} uses a polar coordinate system where the trigger particle defines the $z$-axis. Signals are sought on the cone at a fixed polar angle $\Delta\theta^*$ via the azimuth difference, $\dphi^*$, between two particles. If the away-side jet is aligned with the trigger particle, then Mach-cone signals generated by the away-side jet would appear as a ring of excess particles and result in a constant distribution in $\dphi^*$ with full acceptance. Figure~\ref{fig:phenix_3-particle} shows the background subtracted 3-particle correlation from PHENIX for $1<\pt<2.5<\pttrig<4$~GeV/$c$ in 10-20\% Au+Au collisions. The result is consistent with a Monte Carlo (MC) simulation of conical flow in the PHENIX acceptance, suggesting that the preliminary data are consistent with conical flow~\cite{PHENIX3particle}.

\begin{figure}[hbt]
\hfill
\includegraphics[width=0.90\textwidth,bbllx=0,bblly=15,bburx=565,bbury=320]{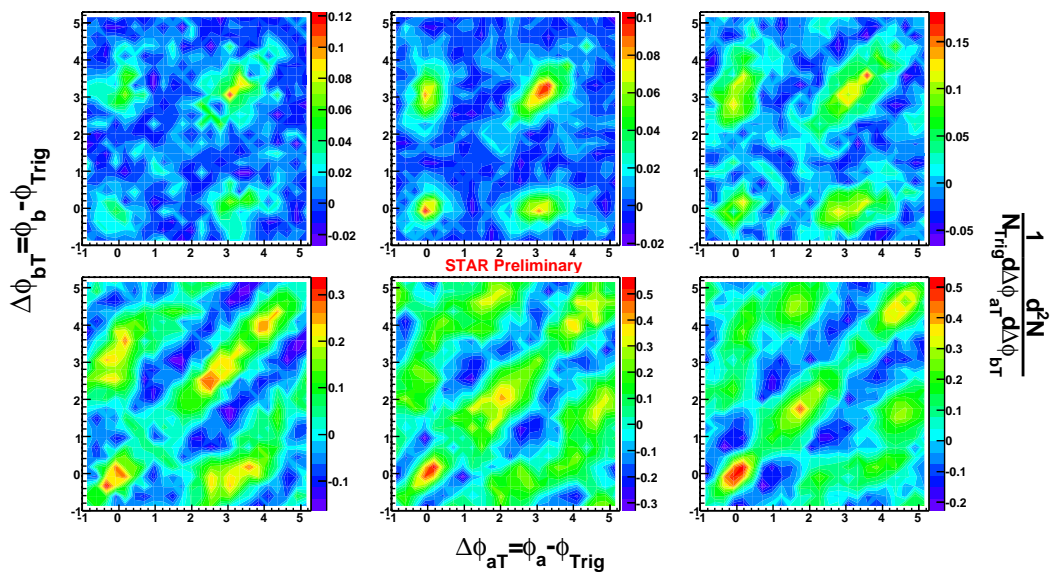}
\caption{(color online) Background subtracted 3-particle jet-like azimuthal correlations from STAR~\cite{STAR3particle} for {\it pp} (top left), d+Au (top middle), and Au+Au 50-80{\%} (top right), 30-50{\%} (bottom left), 10-30{\%} (bottom center), and ZDC triggered 0-12{\%} (bottom right) collisions.}
\label{fig:3-particle}
\end{figure}

STAR~\cite{STAR3particle} analyzes 3-particle jet-like correlations in the azimuthal angles of two associated particles relative to the trigger particle, $\dphi_{aT}$ and $\dphi_{bT}$. Figure~\ref{fig:3-particle} shows the background subtracted 3-particle correlations between a trigger particle with $3<\pttrig<4$ GeV/$c$ and two associated charged particles of $1<\pt<2$ GeV/$c$~\cite{STAR3particle}. The {\it pp}, d+Au and peripheral 50-80\% Au+Au results are similar. Peaks are visible for the near-side, the away-side and the two cases of one particle on the near-side and the other on the away-side.  The peak at ($\pi$,$\pi$) displays a diagonal elongation, consistent with $k_T$ broadening.  The additional broadening in Au+Au may be due to deflected jets.  The more central Au+Au collisions display off-diagonal structure that is consistent with conical emission.  The structure increases in magnitude with centrality and is prominent in the high statistics 12\% central data~\cite{STAR3particle}.

\begin{figure}[hbt]
\hfill
\includegraphics[width=0.85\textwidth,bbllx=0,bblly=0,bburx=560,bbury=180]{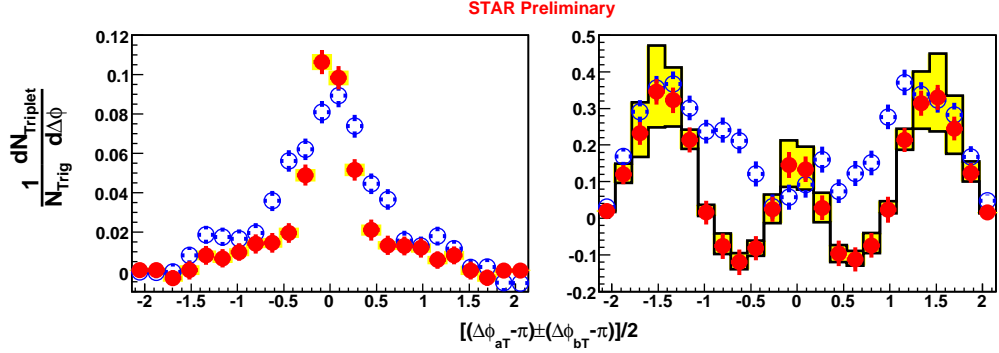}
\vspace{-0.1in}
\caption{(color online) Away-side projections of a strip of width 0.7 radians for d+Au (left) and 0-12\% ZDC triggered Au+Au data (right) from STAR~\cite{STAR3particle}. The off-diagonal projection in $[(\dphi_{aT}-\pi)-(\dphi_{bT}-\pi)]/2$ is shown in solid data points with systematic uncertainties in shaded area, and the on-diagonal projection in $[(\dphi_{aT}-\pi)+(\dphi_{bT}-\pi)]/2$ is shown in open data points.}
\label{fig:projections}
\end{figure}

Figure~\ref{fig:projections} shows the projections of the away-side ($|\dphi_{aT}|>1, |\dphi_{bT}|>1$) 3-particle correlation signal along the on-diagonal ($\dphi_{aT}-\dphi_{bT}=0$) and off-diagonal ($\dphi_{aT}+\dphi_{bT}=2\pi$) axes. Prominent peaks away from $\pi$ are observed in both projections in Au+Au collisions (right panel of Fig.~\ref{fig:projections}). The on-diagonal peaks (open circles) are stronger than the off-diagonal ones (solid circles). 
The off-diagonal peaks are consistent with conical emission. A double Gaussian fit to the projection yields a peak location of 1.45 radians from $\pi$. The on-diagonal peaks may result from the net effect of conical emission and deflected jets. 

%

\begin{figure}[htbp]
\begin{minipage}{0.38\textwidth}
\hspace*{0.4in}
\includegraphics[width=1.2\textwidth,bbllx=0,bblly=20,bburx=520,bbury=375]{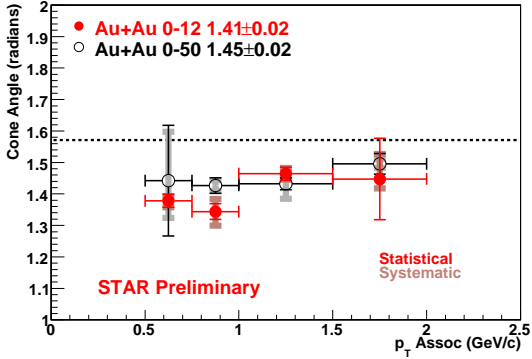}
\end{minipage}
\begin{minipage}{0.6\textwidth}
\caption{(color online) Fitted conical emission angle as a function of the associated particle $\pt$ for 0-12\% ZDC triggered Au+Au (filled) and 0-50\% Au+Au from minimum bias data sample (open) from STAR~\cite{STAR3particle}. Solid error bars are statistical and shaded are systematic. The dashed line is at $\pi/2$. The average cone angles for the two different centralities are indicated in the legend.}
\label{fig:angle_vs_pt}
\end{minipage}
\end{figure}

Conical emission can be generated by Mach-cone shock waves or $\check{\rm C}$erenkov gluon radiation. To distinguish between the two, the $\pt$ dependence of the conical emission angle is studied by STAR, as shown in Figure~\ref{fig:angle_vs_pt}. The observed $\pt$-independent conical emission angle disfavors the simple $\check{\rm C}$erenkov radiation picture. 

If the measured 3-particle correlation structure is indeed due to Mach-cone shock waves, then the conical emission angle gives a direct access to the speed of sound of the medium. Since the system likely evolves through different stages -- the partonic stage quark-gluon plasma, the mixed phase, and the hadronic stage -- to which the initially produced di-jet is sensitive, the measured conical emission is a net effect of the entire evolution history. The extracted speed of sound is, therefore, an ``average" over the evolution of the medium. 

\section{CONCLUSION}

The experimental results on jet-like 2- and 3-particle correlations at RHIC are reviewed. At high transverse momentum, clear away-side correlation peaks are observed that have similar Gaussian widths independent of centrality or $\pt$. Future analysis incorporating a third, soft particle may provide new insights into the underlying physics mechanisms. At low and modest $\pt$, broad, and in some kinematic regions even double-humped, structures are observed on the away side of 2-particle jet-like correlations. The average transverse momentum of the away-side correlated hadrons is larger in the double-hump region than the central region. Mach-cone shock waves have been proposed to explain the observations, however, other physics mechanisms cannot be ruled out without the knowledge of 3-particle jet-like correlation. The 3-particle jet-like correlation results show evidence of conical emission; they also indicate the presence of deflected jets. The $\pt$ independence of the conical emission angle favors conical flow as the underlying physics mechanism. If Mach-cone shock waves are confirmed, further studies can be performed to extract the speed of sound (and the equation of state) of the medium, thereby providing crucial evidence for the creation of quark-gluon plasma at RHIC.

\section*{Acknowledgments}
The author acknowledges discussions with N.~Ajitanand, O.~Barannikova, C.~Gagliardi, J.~Dunlop, J.~Jia, M.~van Leeuwen, B.~Mohanty, C.~Pruneau, S.~Voloshin, J.G.~Ulery. 
This work is supported by U.S. DOE under Grants DE-FG02-02ER41219 and DE-FG02-88ER40412.

\end{document}